\documentclass[10pt]{article}
\usepackage{latexsym}
\usepackage{amssymb}

\usepackage{amsmath}
\usepackage{theorem}

\usepackage{amsfonts,amssymb}

\newtheorem{theorem}{Theorem}
\newtheorem{lemma}{Lemma}

\newtheorem{definition}{Definition}
\newtheorem{remark}{Remark}
\newtheorem{proposition}{Proposition}

\def\be{\begin{equation}}
\def\ee{\end{equation}}
\def\bc{\begin{center}}
\def\ec{\end{center}}

\begin{document}

\title{\bf The mean field Ising model trough interpolating techniques}
\author{ Adriano Barra
  \footnote{King's College London, Department of Mathematics, Strand,
    London WC2R 2LS, United Kingdom and Dipartimento di Fisica,
    Universit\`a di Roma ``La Sapienza'' Piazzale Aldo Moro 2, 00185
    Roma, Italy, {\tt<Adriano.Barra@roma1.infn.it>}}}

\def\be{\begin{equation}}
\def\ee{\end{equation}}
\def\bc{\begin{center}}
\def\ec{\end{center}}

\maketitle

\begin{abstract}
Aim of this work is not trying to explore a macroscopic behavior
of some recent model in statistical mechanics but showing how some
recent techniques developed within the framework of spin glasses
do work on simpler model, focusing on the method and not on the
analyzed system. To fulfil our will the candidate model turns out
to be the paradigmatic mean field Ising model. The model is
introduced and investigated with the interpolation techniques. We
show the existence of the thermodynamic limit, bounds for the free
energy density, the explicit expression for the free energy with
its suitable expansion via the order parameter, the
self-consistency relation, the phase transition, the critical
behavior and the self-averaging properties. At the end a bridge to
a Parisi-like theory is tried and discussed.
\end{abstract}

\section{Introduction}\label{due}

In the past twenty years the statistical mechanics of disordered
systems earned an always increasing weight as a powerful framework
by which analyze the world of complex networks
\cite{amit}\cite{barabasi}\cite{ksat}\cite{contu1}\cite{pagna}.
\newline
The ''harmonic oscillator"  of this field of research is the
Sherrington Kirkpatrick model \cite{MPV} (SK), on which several
schemes have been tested along these years \cite{fischer}; the
first method developed has been the {\itshape replica trick}
\cite{ton3} which, in a nutshell consists in expanding the
logarithm of the partition function $Z(\beta)$ in a power series
of such a function via $\ln Z(\beta) = \lim_{n \rightarrow
0}(Z(\beta)^n-1)/n$, allowing in some way, its analytic
continuation to the $n \rightarrow 0$ limit \cite{MPV}. Such
analytic continuation is not at all simple and many efforts have
been necessary to translate this problem in the language of
theoretical physics built by symmetries and their breaking
\cite{parisi2}. In this scenario a solution has been proposed by
Parisi (and recently proved by Guerra \cite{broken} and Talagrand
\cite{tala}) with the well known Replica Symmetry Breaking scheme,
both solving the SK-model by showing a peculiar ''picture" of the
organization of the underlaying microstructure of this complex
system \cite{mezard}, as well as conferring a key role to the
replica-trick method.
\newline
The replica trick however still pays the price of requiring an ''a
priori" ansatz at some stage of its work and several mathematical
problems concerning its foundations and validity are still open
\cite{T}.
\newline
As a consequence, in the recent past ten years, another method,
called the {\itshape cavity method} \cite{guerra4}, has been
largely improved, mainly thanks to its ability to work without
ansatz and to a natural predisposition for being implemented into
the interpolating technique scheme
\cite{broken}\cite{limterm}\cite{gt2}\cite{barra1}\cite{barra4}.
Also if not so powerful to solve the whole SK-problem without
working in synergy with the replica framework, is, at least for
some kind of questions, a valid alternative to it
\cite{CLT}\cite{gg}\cite{barra5}.
\newline
Aim of this paper is to show some of the features obtainable
within the cavity method by applying it to a simple model, the
mean field Ising model \cite{amit}\cite{parisi2}, which can be
solved with standard methods without requiring nor the replica
trick neither the cavity method itself. Consequently attention
should be payed  on the method, which, once applied on a
paradigmatic model, should be clearer to the non-expert reader
than when applied on complex systems as the SK.
\newline
The paper is structured as follows: Hereafter, still in the first
section, the model is introduced. In section (\ref{tre}) the
interpolating technique for obtaining the thermodynamic limit and
the bounds in the size of the system are discussed. In section
(\ref{quattro}) the interpolating technique to obtain an explicit
expression for the free energy and consequently the phase diagram
are studied. Section (\ref{cinque}) is dedicated to the phase
transition: the lacking of the infinite volume limit against a
vanishing perturbing field, the scaling of the order parameter at
criticality and the self-averaging relations are discussed.  The
last section (\ref{sei}) is a trial introducing technique which
aims to reproduce the Parisi scheme within this simpler framework.

\subsection{Definition of the model and thermodynamics}

The Hamiltonian of the Ising model is defined on $N$ spin
configurations $\sigma:i\to\sigma_i=\pm 1$, labeled by
$i=1,\ldots, N$, as \cite{amit}\cite{parisi2}
\begin{equation}
H_N(\sigma)=-\frac 1N\sum_{1\leq i<j \leq N}\sigma_i \sigma_j
\end{equation}
We assume throughout the paper that there is no external field.
The thermodynamic of the model is carried by the free energy
density $f_N(\beta) =  F_N(\beta)/N$, which is related to the
Hamiltonian via \be e^{-\beta F_N(\beta)} = Z_N(\beta) =
\sum_{\sigma}e^{\frac{\beta} {N}\sum_{1 \leq i<j \leq
N}\sigma_i\sigma_j}, \ee $Z_{N}(\beta)$ being the partition
function.
For the sake of convenience we will not deal with $f_N(\beta)$ but
with the thermodynamic pressure $\alpha(\beta)$ defined via \be
\alpha(\beta) = \lim_{N \rightarrow \infty} \alpha_N (\beta) =
\lim_{N \rightarrow \infty} -\beta f_N(\beta) = \lim_{N
\rightarrow \infty}\frac{1}{N}\ln Z_N(\beta). \ee
A key role will be played by the magnetization $m$, its
fluctuations and its moments, and so let us introduce it as
\be m_N = \frac{1}{N}\sum_{1<i<N} \sigma_i, \ \ \ \langle m_N
\rangle = \frac{\sum_{\sigma} m_N  e^{-\beta
H_N(\sigma)}}{\sum_{\sigma} e^{-\beta H_N(\sigma)}}. \ee
Let us consider also its rescaled fluctuation by introducing the
following random variable
\be\label{csi} \xi_N (\sigma) = \frac{1}{\sqrt{N}}\sum_i\sigma_i
\ee by which the magnetization can be expressed as $\langle m_N
\rangle = \langle \xi_N \rangle N^{-1/2}$; further, let us define
$\gamma(\beta)= 1/ (1-\beta)$  and state, without proof
\cite{ellis}, that in the interval $0<\beta<\beta_c=1$, in the
thermodynamic limit the distribution of $\xi(\sigma) = \lim_{N
\rightarrow \infty}\xi_N(\sigma)$ is a centered Gaussian with
variance equal to $\gamma(\beta)$. The boundary at which the
variance of the distribution diverges (i.e. $\beta=\beta_c=1$)
defines the onset of the broken ergodicity phase.

\section{Thermodynamic limit}\label{tre}

\subsection{Bounding the free energy in the system size}

The first step when dealing with the statistical mechanics package
is, once defined the relevant observable, checking that the model
is well defined (i.e. it admits a good but non trivial
thermodynamic limit). As this task maybe not simple (as for the SK
model or worse for the Hopfield model of neural network
\cite{nota}) working out its $\sup_N$ may help as a first
pre-step. This is usually a simpler task \cite{gallo}. With a
little abuse of language, reminiscent of spin-glass theory, we
call this procedure
 {\em annealing}.
\newline
\newline
\textbf{Annealing of the free energy}
\newline
\newline
Annealed is the thermodynamical regime in which the thermal noise
has a strong effect on the macroscopic behavior of the observable,
while details of the Hamiltonian play a small, thus not
negligible, role. In spin glasses another way to think at the
annealing is by assuming that the dynamics of the spins happens on
the same time-scale of the dynamics of the links between the
spins. For the Ising model there is no true annealing as there are
no quenched variables and the closest procedure to be performed
can be obtained trivially as follows:
\begin{eqnarray} Z_N(\beta) &\leq&
\sum_{\sigma}e^{\beta/N}e^{\frac{N(N-1)}{2}} \leq 2^N
e^{\frac{\beta (N-1)}{2}} \\ \label{annealing} \frac{1}{N}\ln
Z_N(\beta) &\leq& \ln 2 + \frac{\beta}{2}(1 - \frac{1}{N})
\Rightarrow \alpha(\beta) \leq \ln 2 + \frac{\beta}{2}
\end{eqnarray}
Following this approach the next step is trying and bound, in the
volume size, the free energy from above and from below. For the
Ising model this can be obtained as follows:
\newline
\newline
\textbf{Upper bound of the free energy}
\newline
\newline
While for disordered systems bounding the free energy in the
volume limit is not an easy task, for model with no disorder such
bounds can be easily obtained  \cite{luca}\cite{leshouches}.
Consider the trivial estimate of the magnetization $m$, valid for
all trial fixed magnetization $M$
\begin{equation}
\label{ } m^2 \geq 2mM-M^2
\end{equation}
and plug it into the partition function to get (neglecting terms
vanishing in the thermodynamic limit)
\begin{equation*}
Z_N(\beta) = \sum_{\sigma}e^{\frac{\beta}{N}\sum_{1\leq i<j \leq
N}\sigma_i\sigma_j}= \sum_{\sigma}e^{\frac{\beta N m^2}{2}} \geq
\sum_\sigma e^{\beta m M N}e^{-\frac12 \beta J M^2 N}.
\end{equation*}
Now this sum is easy to compute, since the magnetization appears
linearly and therefore the sum factorizes in each spin. Physically
speaking, we replaced the two-body interaction, which is difficult
to deal with, with a one-body interaction. Then we try to
compensate this by modulating the field acting on each spin by
means of a trial fixed magnetization and a correction term
quadratic in this trial magnetization $M$.
\begin{remark}
This idea is reminiscent of a recent powerful method
\cite{ass}\cite{barra2}\cite{g} introduced by Aizenman and
coworkers for the spin-glass theory, in which the key idea, is
letting interact the system one is dealing with, with an external
structure in such a way by which, sending the size of the this
structure to infinity, thanks to the mean field nature of the
interaction, the system no longer interacts with itself, making
the mathematical control simpler.
\end{remark}
The result is the following bound
\begin{equation}\label{cwbound1}
\frac1N \ln Z_N(\beta)\geq \sup_{M}\{\ln 2+\ln\cosh(\beta  M) -
\frac12 \beta  M^2\}
\end{equation}
that holds for any size of the system $N$. The result is quite
typical, the term $\ln 2$ is there because the sum over a spin of
a Boltzmann factor linear in the spins is {\sl twice} the
hyperbolic cosine, which appears as second term (that essentially
gives the entropy). The third term is the internal energy
(multiplied by $-\beta$).
\newline
\newline
\textbf{Lower bound of the free energy}
\newline
\newline
In order to get the opposite bound to (\ref{cwbound1}), let us
notice that the magnetization $m$ can take only $2N+1$ distinct
values. We can therefore split the partition function into sums
over configurations with constant magnetization in the following
way
\begin{equation}
Z_N(\beta)=\sum_\sigma\sum_M\delta_{mM}e^{\frac12 \beta  Nm^2}
\end{equation}
using the trivial identity
\begin{equation}
\sum_M\delta_{mM}=1\ .
\end{equation}
Now inside the sum $m=M$, which means also
\begin{equation}\label{fix}
m^2=2mM-M^2\ .
\end{equation}
Plugging the latter equality into $Z_N(\beta)$ and using the
trivial inequality
$$\label{}
\delta_{mM}\leq 1
$$
yields
\begin{equation}
Z_N(\beta)\leq\sum_M\sum_\sigma e^{\beta  NmM}e^{-\frac12 \beta N
M^2}\ .
\end{equation}
Now one can carry out the sum over $\sigma$ bounding the remaining
sum over $M$ by $2N+1$ times its largest term gives then
\begin{equation}
Z_N(\beta)\leq\sum_M\sup_M\{\ln 2+\ln\cosh(\beta M) - \frac12
\beta  M^2\}
\end{equation}
from which
\begin{equation}\label{cwbound2}
\frac1N \ln Z_N(\beta)\leq \ln \frac{2N+1}{N}+\sup_M\{\ln
2+\ln\cosh(\beta  M) - \frac12 \beta  M^2\}\ .
\end{equation}
This gives, together with (\ref{cwbound1}), the exact value of
free energy per site at least in the thermodynamic limit.

\subsection{Bound by interpolating the size of the system}\label{guerratoninelli}

A breakthrough in showing the existence of the thermodynamic limit
for mean field disordered systems has been obtained recently
within the Guerra-Toninelli interpolation scheme \cite{limterm}.
Previously several beautiful model-specific attempts were made
\cite{comets}\cite{bovier}\cite{boviergrem}, but this
interpolating scheme showed an immediate wide range of
applications and its beauty is its simplicity. We are going to
introduce it applied to the Ising-model.
\newline
Divide the $N$ spin system into two subsystems of $N_1$ and $N_2$
spins each, with $N_1+N_2=N$. Denoting by $m_{1}(\sigma)$,
$m_2(\sigma)$ the magnetization corresponding to the subsystems,
{\it i.e.}
\begin{equation*}
m_1(\sigma)=\frac1{N_1}\sum_{i=1}^{N_1}\sigma_i\ ,\
m_2(\sigma)=\frac1{N_2}\sum_{i=N_1+1}^{N}\sigma_i,
\end{equation*}
one sees that $m(\sigma)$ is a convex linear combination of
$m_1(\sigma)$ and $m_2(\sigma)$:
\begin{equation}
m(\sigma)=\frac{N_1}Nm_1(\sigma)+\frac{N_2}Nm_2(\sigma).
\end{equation}
Since the function $x\to x^2$ is convex, one has
\begin{equation*}
Z_N(\beta)\leq \sum_{\{\sigma\}}\exp (\beta (N_1 m_1^2(\sigma)+N_2
m_2^2(\sigma))) =Z_{N_1}(\beta)Z_{N_2}(\beta)
\end{equation*}
and
\begin{equation}
\label{arisuper} N f_N(\beta)=-\frac1\beta\ln Z_N(\beta)\ge N_1
f_{N_1}(\beta)+N_2 f_{N_2}(\beta).
\end{equation}
\begin{theorem}
The infinite volume limit for $\alpha_N(\beta)$ does exist and
equals its sup. \be \lim_{N \rightarrow \infty} \alpha_N(\beta) =
\sup_N \alpha_N(\beta) \equiv \alpha(\beta) \ee
\end{theorem}
\textbf{Proof}
\newline
In a nutshell the two key ingredients are the subadditivity
 $(Nf_N \geq N_1f_1 + N_2f_2)$ and the property
of the free energy density of being limited from above uniformly
in $N$ which is established elementary by using the annealing. It
is also evident by considering Eq.s
(\ref{cwbound1},\ref{cwbound2}) $\Box$.
\newline
Unfortunately, the very simple approach we illustrated above as it
is, does not apply to the SK model, where the randomness of the
couplings prevents us from exploiting subadditivity directly on
the Hamiltonian $H_N$. However, the related strategy, which allows
in some sense an extension to mean field spin glass models is to
interpolate between the original systems of $N$ spins, and two
non-interacting systems, containing $N_1$ and $N_2$ spins,
respectively, and to compare the corresponding free energies. To
this purpose, consider the interpolating parameter $0\le t\le1$,
and the auxiliary partition function
\begin{equation}
\label{zntCW} Z_N(t)=\sum_{\{\sigma\}}\exp(\beta\left(N t
m^2(\sigma)+N_1(1-t)m_1^2(\sigma) +N_2(1-t)m_2^2(\sigma)\right)).
\end{equation}
Of course, for the boundary values $t=0,1$ one has
\begin{eqnarray}
\label{aribounds} &&-\frac1{N\beta}\ln Z_N(1)=f_N(\beta)\\
 \label{ari2} &&-\frac1{N\beta}\ln
Z_N(0)=\frac{N_1}Nf_{N_1}(\beta)+\frac{N_2}N f_{N_2}(\beta)
\end{eqnarray}
and, taking the derivative with respect to $t$,
\begin{equation}
-\frac d{dt}\frac1{N\beta}\ln Z_N(t)=- \left\langle
m^2(\sigma)-\frac{N_1}N
m_1^2(\sigma)-\frac{N_2}Nm_2^2(\sigma)\right\rangle_t \ge0,
\end{equation}
where $\langle \ \rangle_t$ denotes  the Boltzmann-Gibbs thermal
average with the extended weight encoded in the $t$-dependent
partition function (\ref{zntCW}). Therefore, integrating in $t$
between $0$ and $1$, and recalling the boundary conditions
(\ref{aribounds},$20$), one finds again the superadditivity
property (\ref{arisuper}).

The interpolation method, which may look unnecessarily complicated
for the Curie-Weiss model, is actually the only one working in the
case of mean field spin glass systems.

\section{The structure of the free energy}\label{quattro}

In this chapter we adapt the work \cite{barra1} developed for the
SK model to the mean field Ising model.
\newline
The main idea of the {\em cavity field} method is to look for an
explicit expression of $\alpha_N(\beta)=-\beta f_N(\beta)$ upon
increasing the size of the system from $N$ particles (the cavity)
to $N+1$ so that, in the limit of N that goes to infinity
\cite{Glocarno}\cite{guerra2}

\be\lim_{N \rightarrow \infty}\frac{(-\beta
F_{N+1}(\beta))-(-\beta F_N(\beta))}{(N+1-N)}=-\beta f(\beta)\ee
because the existence of the thermodynamic limit (sec.
\ref{guerratoninelli}) implies only vanishing correction of the
free energy density.

\subsection{Interpolating cavity field}

As we will see, the interpolating technique can be very naturally
implemented in the cavity method; let us consider the partition
function of a system made by $N+1$ spins: \be Z_{N+1}(\beta) =
\sum_{\sigma}e^{-\beta H_{N+1}(\sigma)} = \sum_{\sigma_{N+1}= \pm
1}\sum_{\sigma} e^{\frac{\beta}{N+1}
\sum_{1<i<j<N}\sigma_i\sigma_j}e^{\frac{\beta}{N+1}
\sum_{1<i<N}\sigma_i\sigma_{N+1}}. \ee
With the gauge transformation $\sigma_i \rightarrow
\sigma_i\sigma_{N+1}$, which, of course, is a symmetry of the
Hamiltonian, we get
\be Z_{N+1}\label{049} (\beta) = 2 Z_N(\beta^*)
\tilde{\omega}(e^{\frac{\beta}{N+1}\sum_{1<i<N}\sigma_i}) \ee
where $\tilde{\omega}$ is the Boltzmann state at the inverse
temperature $\beta^* = \beta \frac{N}{N+1}$ (note that in the
thermodynamic limit the shifted temperature converges to the real
one $\beta^* \rightarrow \beta$).
Let us reverse the temperature shift and apply the logarithm to
both the sides of Eq. (\ref{049}) to obtain
\be\label{051} \ln Z_{N+1}(\beta \frac{N+1}{N}) = \ln 2 + \ln
Z_N(\beta) + \ln \omega_N(e^{\frac{\beta}{N}\sum_{1<i<N}\sigma_i})
\ee
Equation (\ref{051}) tell us that via the third term of  its
r.h.s. we can bridge an Ising system with $N$ particles at an
inverse temperature $\beta$ to an Ising system with $N+1$
particles at a shifted inverse temperature $\beta^* = \beta
(N+1)/N$. Focusing on such a term let us make the following
definitions.
\begin{definition}
We define an extended partition function $Z_N(\beta,t)$ as
\be\label{partiz} Z_N(\beta,t) = \sum_{\sigma}e^{-\beta
H_N(\sigma)}e^{\frac{t}{N}\sum_{1<i<N}\sigma_i} \ee
\end{definition}

Note that the above partition function, at $t=\beta$, turns out to
be, via the global gauge symmetry $\sigma_i \rightarrow
\sigma_i\sigma_{N+1}$, a partition function for a system of $N+1$
spins at a shifted temperature $\beta^*$  apart a constant term.
On the same line
\begin{definition}
we define the generalized Boltzmann state $\langle \ \rangle_t$ as
\be \langle F(\sigma) \rangle_t = \frac{\langle F(\sigma)
e^{\frac{t}{N}\sum_{1<i<N}\sigma_i} \rangle }{ \langle
e^{\frac{t}{N}\sum_{1<i<N}\sigma_i  \rangle}}, \ee $F(\sigma)$
being a generic function of the spins.
\end{definition}

\begin{definition} Related to the Boltzmann state $\langle \ \rangle$
we define the cavity function $\Psi(\beta,t)=\lim_{N \rightarrow
\infty}\Psi_N(\beta,t)$ as \be\label{052} \Psi_N(\beta,t) = \ln
\langle e^{\frac{t}{N}\sum_{1<i<N} \sigma_i}\rangle \ee
\end{definition}

\begin{proposition} The cavity function $\Psi(\beta,t)$ is the
generating function of the centered momenta of the magnetization,
examples of which are
\begin{eqnarray}\label{generatrice} \frac{\partial \Psi_N(\beta,t)}{\partial t}
&=& \langle m_N \rangle_t \\  \frac{\partial^2
\Psi_N(\beta,t)}{\partial t^2} &=& \langle m^2_N \rangle_t -
\langle m_N \rangle^2_t
\end{eqnarray}
\end{proposition}
\textbf{Proof}
\newline
The proof is straightforward and can be obtained by simple
derivation:
\begin{eqnarray} \nonumber
\frac{\partial \Psi_N(\beta,t)}{\partial t} &=& \partial_t \ln
\omega_N(e^{\frac{t}{N}\sum_{1<i<N}\sigma_i}) = \partial_t \ln
\sum_{\sigma} e^{-\beta H_N(\sigma)}e^{\frac{t}{N}\sum_i \sigma_i}
= \\ \nonumber &=& \frac{\sum_{\sigma} \frac{1}{N} \sum_{1<i<N}
\sigma_i e^{-\beta H_N(\sigma)}e^{\frac{t}{N}\sum_i
\sigma_i}}{\sum_{\sigma} e^{-\beta
H_N(\sigma)}e^{\frac{t}{N}\sum_i \sigma_i}} = \langle m_N
\rangle_t
\end{eqnarray}
The second derivative is worked out exactly as the first $\Box$.
\begin{remark}
We stress that in the disordered counterpart (i.e. the SK model) a
proper interpolating cavity function is defined by introducing
$\sqrt{t}$ instead of $t$. This reflects the property of the
Gaussian coupling of adding another extra derivation due to Wick
theorem. It is worth nothing that again the Gaussian coupling
makes necessary the normalization factor $\sqrt{N}$ instead of $N$
in front of the Hamiltonian such that the adaptation from Ising
$t/N$ to SK $\sqrt{t/N}$ is the same for $t$ and $N$.
\end{remark}
\begin{definition}
We define respectively as fillable and filled monomials the odd
and even momenta of the magnetization weighted by the extended
Boltzmann measure such that
\begin{itemize}

\item $\langle m_N^{2n+1} \rangle_t$ with $n \in \mathbb{N}$ is
fillable

\item  $\langle m_N^{2n} \rangle_t$ with $n \in \mathbb{N}$ is
filled

\end{itemize}
\end{definition}

\subsection{Saturability and gauge-invariance}

The next step  is to motivate why we introduced the whole
machinery: The first reason we are going to show are peculiar
properties of both the filled and the fillable monomials. In the
thermodynamic limit, the first class do not depend on the
perturbation induced by the cavity field and, at $t=\beta$, the
latter (via the $\sigma_i \rightarrow \sigma_i\sigma_{N+1}$
symmetry) is projected into the first class. The second reason is
that the free energy can be expanded via these monomials, so a
good control of them means a good knowledge of the thermodynamic
of the system.

\begin{theorem}
In the $N \rightarrow \infty$ limit  the averages $\langle
m_N^{2n} \rangle$ of the filled monomials are $t$-independent for
almost all values of $\beta$, such that
\end{theorem}
$$
\lim_{N \rightarrow \infty} \partial_t \langle m_N^{2n}  \rangle_t
= 0
$$
\textbf{Proof}
\newline
Without loss of generality we will prove the theorem in the
simplest case (for $\langle m_N^{2}  \rangle$); it will appear
immediately clear how to generalize the proof to higher order
monomials. Let us write the cavity function as \be\label{cf13}
\Psi_N(\beta,t)=\ln Z_{N}(\beta,t)-\ln Z_{N}(\beta) \ee
and derive it with respect to $\beta$: \be \label{cf16}
\frac{\partial \Psi_N(\beta,t)}{ \partial \beta}=\frac {N}{2}(
\langle m_N^{2} \rangle - \langle m_N^{2} \rangle_t). \ee We can
introduce an auxiliary function $\Upsilon_N(\beta,t)= (\langle
m_N^{2} \rangle - \langle m_N^{2}  \rangle_t)$ such that: \be
\Upsilon_N(\beta,t)=\frac{2}{N}\partial_{\beta} \Psi_N(\beta,t)
\ee
and integrate it in a generic interval $[\beta_1,\beta_2]$:
\be\label{contesto}
\int_{\beta_1}^{\beta_2}\Upsilon_N(\beta,t)d\beta^2=
\frac{4}{N}[\Psi_N(\beta_2,t)-\Psi_N(\beta_1,t)]. \ee Now we must
control $\Psi_N(\beta,t)$ in the $N \rightarrow \infty$ limit; the
simplest way is to look at its $t$-streaming $\partial_t
\Psi_N(\beta,t) = \langle m_N \rangle_t $ such   the
$N$-dependence is just taken into account by the Boltzmann factor
inside the averages and, as $\langle m_N \rangle_t \in [-1,1]$, in
the thermodynamic limit $\Psi(\beta,t)$ remains bounded and the
second member of (\ref{contesto})  goes to zero such that,
$\forall$ [$\beta_1$,$\beta_2$], $\Upsilon_N(\beta,t)$ converges
to zero implying $\langle m_N^2 \rangle_t \rightarrow \langle
m_N^2 \rangle$$\Box$.
\begin{remark}A consequence of this property, in the spin glass theory,
turns out to be the stochastic stability
\cite{parisiSS}\cite{cg2}.
\end{remark}
The next theorem is crucial for this section, so, for the sake of
simplicity, we split it in two part: at first we prove  the
following lemma than it will make us able to proof the core of the
theorem itself which will be showed immediately after. For a
clearer statement of the lemma we take the freedom of pasting the
volume dependence of the averages as a subscript close to the
perturbing tuning parameter $t$.

\begin{lemma}\label{pigreco} Let  $\langle \ \rangle_N$ and $\langle \ \rangle_{N,t}$ be the
states defined, on a system of $N$ spins, respectively by the
canonical partition function $Z_N(\beta)$ and by the extended one
$Z_N(\beta,t)$; if we consider the ensemble of indexes
$\{i_1,..,i_r\}$ with $r \in [1,N]$, then for $t=\beta$, where the
two measures become comparable, thanks to the global gauge
symmetry (i.e. the substitution $\sigma_i \rightarrow
\sigma_i\sigma_{N+1}$) the following relation holds \be
\omega_{N,t=\beta}(\sigma_{i_1}...\sigma_{i_r})=
\omega_{N+1}(\sigma_{i_1}...\sigma_{i_r}\sigma_{N+1}^r)
+O(\frac{1}{N} ) \ee where r is an exponent, not a replica index,
so if r is even $\sigma_{N+1}^r=1$, while if it is odd
$\sigma_{N+1}^r=\sigma_{N+1}$.
\end{lemma}
\textbf{Proof}
\newline
Let us write $\omega_{N,t}$ for $t=\beta$, defining for the sake
of simplicity $\pi=\sigma_{i_1}...\sigma_{i_r}$:
\be \omega_{N,t=\beta}(\sigma)=
[\sum_{{\sigma}}\frac{1}{Z_{N}(\beta)}
e^{\frac{\beta}{\sqrt{N}}\sum_{1 \leq i<j \leq
N}J_{ij}\sigma_i\sigma_j + \frac{\beta}{\sqrt{N}}\sum_i
J_i\sigma_i}\pi]. \ee
Introducing first a sum over $\sigma_{N+1}$ at the numerator and
at the denominator, (which is the same as multiply and divide for
$2^N$ because there is still no dependence to $\sigma_{N+1}$) and
making the transformation  $\sigma_i\rightarrow
\sigma_i\sigma_{N+1}$, the variable $\sigma_{N+1}$ appears at the
numerator and it is possible to build the status at $N+1$
particles with the little temperature shift which vanishes in the
thermodynamic limit:
\be \omega_{N,t= \beta}(\sigma)=
\omega_{N+1}(\sigma\sigma_{N+1}^r) +O(\frac{1}{N})\Box. \ee
Using this lemma we are able to proof the following
\begin{theorem}\label{saturi}
{\itshape Let $\langle M \rangle$ be a fillable monomial of the
magnetization, (this means that  \ $\langle mM \rangle$ is
filled). We have:}
\be \lim_{N \rightarrow \infty}\lim_{t\rightarrow\beta} \langle M
\rangle_t= \langle mM \rangle
\ee
\end{theorem}
\textbf{Proof}
\newline
The proof is a straightforward application of Lemma
\ref{pigreco}$\Box$.
\newline

\subsection{The free energy via the interpolating cavity method}

The fact that the free energy is expressed as the difference
between an entropy term coming from a one-body interaction and the
internal energy times $\beta$ is typical of thermodynamics. We
found this feature when looking at the bounds
(\ref{cwbound1}),(\ref{cwbound2}); now, stating the next
fundamental theorem,  we find the same structure via this
interpolating version of the cavity field method.

\begin{theorem}
The following relation holds in the thermodynamic limit:
\be\label{215} \alpha(\beta) = \ln 2+\Psi(t=\beta) - \beta
\frac{\partial \alpha (\beta)}{\partial \beta} \ee
\end{theorem}
\textbf{Proof}
\newline
Let us consider again the partition function of a system made up
by $(N+1)$ spins and point out with $\beta$ the true temperature
and with $\beta^* = \beta(1 + N^{-1})$ the shifted one:
\be\label{217} Z_{N+1}(\beta) =\sum_{{\sigma_{N+1}}}
e^{\frac{\beta}{\sqrt{N+1}}\Sigma_{1 \leq i<j \leq
N+1}\sigma_i\sigma_j}=
 2\sum_{{\sigma_{N}}}
e^{\frac{\beta^*}{\sqrt{N}}\Sigma_{1 \leq i<j \leq
N}\sigma_i\sigma_j} e^{\frac{\beta}{\sqrt{N+1}}\sum_{1<i<N}
\sigma_i}, \ee
Now we multiply and divide by $Z_N(\beta^*)$ the right hand side
of eq. \ (\ref{217}), then we take the logarithm on both sides and
 subtract from every member the quantity $\ln Z_{N+1}(\beta^*)$;
expanding $\ln Z_{N+1}(\beta)$ around $\beta= \beta^*$ as \be \ln
Z_{N+1}(\beta)- \ln Z_{N+1}(\beta^*) = (\beta- \beta^*)
\partial_{\beta^*} \ln Z_{N+1}(\beta^*)+ O((\beta-\beta^*)^2)
\ee
with
\be
\beta-\beta^*=\beta^*(\sqrt{\frac{N+1}{N}}-1)=\frac{\beta^*}{2N} +
O(N^{-1}) \ee
we substitute $\beta$ with $\beta^*$ inside the state $\omega$ and
neglecting corrections $O(N^{-1})$ we have:
\begin{eqnarray}\nonumber && \ln Z_{N+1}(\beta^*) +(\beta-\beta^*)\partial_{\beta^*}\ln
Z_{N+1}(\beta^*)= \\ && \ln 2+\ln Z_N(\beta^*)+ \ln \omega_{N,
\beta^*}(e^{\frac{\beta}{\sqrt{N+1}} \sum_{1<i<N} J_i\sigma_i})
+O(N^{-1}), \end{eqnarray}
where, with the symbol $\omega_{N,\beta^*}$ we stressed that the
temperature inside the Bolzmann average is the shifted one. Using
the variable $\alpha(\beta^*)$ and renaming $\beta^* \rightarrow
\beta$ in the thermodynamic limit we get: \be \alpha(\beta) +
\beta\frac{d \alpha(\beta)}{d \beta} = \ln 2+ \Psi(t=\beta). \ee
and this is the thesis of the theorem $\Box$.
\newline

\subsection{Self-consistency of the order parameter via its streaming}

As we saw in the last section the cavity function is deeply
related to the free energy. Usually the internal energy is much
simpler to evaluate than the free energy because there is no
contribution by the entropy, which, especially in complex system,
can make things much harder; consequently if we learn how to
extrapolate information from the cavity function we can obtain
information for the free energy. To fulfil this task we state the
following theorem.
\begin{theorem}
When taken a generic well defined function of the spins
$F(\sigma)$, the following streaming equation holds:
\be\label{streaming} \frac{\partial \langle F_N(\sigma)
\rangle_t}{\partial t} = \langle F_N(\sigma) m_N \rangle_t -
\langle F_N(\sigma) \rangle_t \langle m_N \rangle_t \ee
\end{theorem}
\textbf{Proof}
\newline
The proof is straightforward and can be obtained by simple
derivation:
\begin{eqnarray}\nonumber &&
\frac{\partial \langle F_N(\sigma) \rangle_t}{ \partial t} =
\partial_t \frac{\sum_{\sigma}F_N(\sigma)e^{-\beta H_N(\sigma)} e^{\frac{t}{N}\sum_{1<i<N}\sigma_i}}
{\sum_{\sigma}e^{-\beta H_N(\sigma)}
e^{\frac{t}{N}\sum_{1<i<N}\sigma_i}}  \\ \nonumber &&= \Big(
\frac{\sum_{\sigma}F_N(\sigma) \frac{1}{N}\sum_{1<i<N} \sigma_i
e^{-\beta H_N(\sigma)} e^{\frac{t}{N}\sum_{1<i<N}\sigma_i}}
{\sum_{\sigma}e^{-\beta H_N(\sigma)}} \Big) - \\ \nonumber &&\Big(
\frac{\sum_{\sigma}F_N(\sigma) e^{-\beta H_N(\sigma)}
e^{\frac{t}{N}\sum_{1<i<N}\sigma_i}} {\sum_{\sigma}e^{-\beta
H_N(\sigma)}} \Big) \times \\ \nonumber && \times  \Big(
\frac{\sum_{\sigma} \frac{1}{N}\sum_{1<i<N} \sigma_i e^{-\beta
H_N(\sigma)} e^{\frac{t}{N}\sum_{1<i<N}\sigma_i}}
{\sum_{\sigma}e^{-\beta H_N(\sigma)}} \Big)  \\ \nonumber
&&=\langle  F_N(\sigma)m_N \rangle_t - \langle F_N(\sigma)
\rangle_t \langle m_N \rangle_t.\Box
\end{eqnarray}
We now want to expand via filled monomials of the magnetization
the cavity function by applying the streaming equation
(\ref{streaming}) directly to its derivative, thanks to Eq.
(\ref{generatrice}). It is immediate to find that the streaming of
$\langle m_N \rangle_t$ obeys the following differential equation
\be\label{riusa}
\partial_t \langle m_N \rangle_t = \langle m^2_N \rangle_t -
\langle m_N \rangle_t^2 \ee which,  thanks to Theorem
(\ref{saturi}), becomes trivial in the thermodynamic limit. In
fact, calling $m = \lim_{N \rightarrow \infty} m_N$ and skipping
the subscript $t$ on $\lim_{N \rightarrow \infty}\langle m_N^2
\rangle_t = \langle m^2 \rangle$ we obtain
$$
\frac{1}{\langle m^2 \rangle}\partial_t \langle m \rangle_t = 1 -
(\frac{\langle m \rangle_t^2}{\langle m^2 \rangle})
$$
which is easily solved by splitting the variables and the solution
is
\be\label{semitgh} \langle m \rangle_t = \sqrt{\langle m^2
\rangle} \tanh (\sqrt{\langle m^2 \rangle} t). \ee
Once evaluated  Eq. (\ref{semitgh}) by using the gauge at
$t=\beta$ (i.e. $\langle m \rangle_{t=\beta} = \langle m^2 \rangle
$) we get
\be\label{tgh} \sqrt{\langle m^2 \rangle} = \tanh(\beta
\sqrt{\langle m^2 \rangle}) \ee
which is the well known self-consistency equation for the
Ising-model.

\subsection{The free energy expansion}

From Eq. (\ref{semitgh}) it is possible to obtain an explicit
expression for the cavity function to plug into Eq.(\ref{215})
solving for the free energy. In fact we have
\be \lim_{N \rightarrow \infty}\Psi_N(\beta,t) = \lim_{N
\rightarrow \infty} \int_0^t dt' \langle m_N \rangle_{t'} =
\int_0^t dt' \sqrt{\langle m^2 \rangle} \tanh (\sqrt{\langle m^2
\rangle} t)  \ee from which is immediate to solve for the
$\Psi(\beta,t)$:
\be\label{compegno} \Psi(\beta,t) = \ln\cosh{(\sqrt{\langle m^2
\rangle} t)}. \ee
The last term still missing to fulfil the expression of the free
energy via eq.(\ref{215}), which is immediate to obtain, is the
internal energy.
\begin{proposition}
The internal energy of the Ising model is
\be \beta \frac{d \alpha_N(\beta)}{d \beta} =
\frac{\beta}{2}\langle m_N^2 \rangle\ee
\end{proposition}
\textbf{Proof}
\newline
The proof is straightforward and can be obtained by simple
derivation on the same line of the previous proofs $\Box$.
\newline
\newline
Pasting all together we have
\begin{proposition}\label{busillis}
The free energy of the Ising model is \be\label{freeE}
\alpha(\beta) = \ln 2 + \ln \cosh(\beta \sqrt{\langle m^2
\rangle}) - \frac{\beta}{2} \Big( \sqrt{\langle m^2 \rangle}
\Big)^2 \ee
\end{proposition}
\textbf{Proof}
\newline
The proof proceeds by making explicit Eq.(\ref{215}). $\Box$

\section{The phase transition}\label{cinque}

\subsection{Breaking commutativity of volume and vanishing perturbation limit
}\label{break}

The reasoning of this section can be found, always in the context
of spin glasses in \cite{barra6}.
\newline
Let us move one step backward and consider Eq. (\ref{freeE}) at
finite $N$. The receipt to obtain the expression of the free
energy via the filled monomial is to perform at first the
$N\rightarrow \infty$ limit to saturate the fillable term and then
the $t \rightarrow \beta$ limit to free the measure from the
perturbation (making it works as a cavity field). So in other
words $\alpha(\beta) = \lim_{t \rightarrow \beta}\lim_{N
\rightarrow \infty}\alpha_N(\beta,t)$. But what if we exchange the
limits such that $\alpha^*(\beta) = \lim_{N \rightarrow
\infty}\lim_{t \rightarrow \beta}\alpha_N(\beta,t)$?
\newline
Simply, thanks to the gauge invariance $\lim_{N \rightarrow
\infty}\lim_{t \rightarrow \beta}\langle m_N \rangle = 0$ implying
$\Psi(\beta,t)=0$, defining the high temperature expression for
$\alpha^*(\beta)$.
\newline
Alternatively one can solve Eq. (\ref{riusa}) for the variable
$\langle \xi_N(\sigma) \rangle_t$ by sending first $N \rightarrow
\infty$ and check that these fluctuations scale accordingly the
paragraph after Eq. (\ref{csi}).
\newline
Coherently  there is a range in temperature (the paramagnetic
phase) in which $\alpha(\beta) = \alpha^*(\beta)$ such that the
two limits $\lim_{t \rightarrow \beta}\lim_{N \rightarrow \infty}$
  do
commute. This can be understood as follows: If we consider just
the ``high temperature region" saturability implies $\langle m^2
\rangle = 0$ (because $\lim_{N \rightarrow \infty}\langle m
\rangle_t \rightarrow \langle m^2 \rangle \in [0,1]$ such that
$\langle m_N^2 \rangle = 0,1$ but $\langle m^2 (\beta=0) \rangle =
0$) and the high temperature expression holds. In the range $\beta
\in [0,1]$ the global symmetry of the Hamiltonian $\sigma_i
\rightarrow \sigma_i\sigma_{N+1}$ is a symmetry of the Boltzmann
state too, while in the range $\beta \in \ ]1,\infty]$ the
Boltzmann state shares no longer this invariance and ergodicity is
lost. In the next section the finding of such a critical point,
which defines the onset of ergodicity breaking, is discussed
together with the control of the system at criticality.
\newline

\subsection{Critical behavior: scaling laws}

Critical exponents are needed to characterize singularities of the
theory at the critical point and, for us, this information is
encoded in the behavior of the order parameter $\sqrt{\langle m^2
\rangle}$.
\newline
Assuming for the moment that $\beta_c=1$ (where $\beta_c$ stands
for the critical point in temperature), close to criticality, we
take the freedom of writing $G(\beta) \sim G_0 \cdot
(\beta-1)^{\gamma}$, where the symbol $\sim$ has the meaning that
the term at the second member is the dominant but there are
corrections of order higher than $\tau^{\gamma}$.
\newline
The standard way to look at the scaling of the order parameter is
by expanding the hyperbolic tangent around $\sqrt{\langle m^2
\rangle} \sim 0$ obtaining
\be \sqrt{\langle m^2 \rangle} = \tanh(\beta \sqrt{\langle m^2
\rangle}) \sim \beta \sqrt{\langle m^2 \rangle} - \frac{(\beta
\sqrt{\langle m^2 \rangle)}^3}{3} \ee
by which one gets  \be\label{dicastro} \sqrt{\langle m^2
\rangle}(1 - \beta) + \frac13 (\beta (\sqrt{\langle m^2
\rangle})^3) \sim 0. \ee The first solution of eq.(\ref{dicastro})
is $\sqrt{\langle m^2 \rangle}=0$ (which is also the only solution
in the ergodic phase) while the other two solutions can be
obtained by solving \be (\sqrt{\langle m^2 \rangle})^2 \sim
\frac{(\beta-1)^3}{\beta^3} \sim 3(1 - \frac1\beta)\ee  close to
the critical point, obtaining
\be\label{prova} \sqrt{\langle m^2 \rangle} \sim
(\beta-1)^{\frac12} \ee which gives as the critical exponent
$\gamma = 1/2$.
\newline
\newline
Within our framework the procedure is by using directly the
streaming equation (\ref{streaming}), expanding iteratively in
filled monomials, obtaining
\begin{eqnarray}\label{espa}
\langle m \rangle_t &=& \langle m^2 \rangle t - \int_0^t \langle m
\rangle^2_t  \\ \nonumber &=& \langle m^2 \rangle t - \int_0^t dt'
\Big( \langle m^2 \rangle^2 t'^2 -2 \langle m^2 \rangle t'
\int_0^{t'} dt'' \langle m \rangle^2_{t''} + ( \int_0^{t'} dt''
\langle m \rangle^2_{t''} ) \Big) \\ \nonumber &=& \langle m^2
\rangle t - \langle m^2 \rangle^2 \frac{t^3}{3} + O(\langle m^2
\rangle^4),
\end{eqnarray}
where higher order terms, close to criticality, can be neglected.
Now by applying saturability (Theorem \ref{saturi}) at $t=\beta$
we get \be\label{scaling} \langle m^2 \rangle (\beta-1) = \langle
m^2 \rangle^2 \frac{\beta^3}{3} + O(\langle m^2 \rangle^4)\ee from
which we can derive both the critical point and the scaling
exponent: To find the critical point it is enough to rewrite
eq.(\ref{scaling}) switching to the rescaled order parameter
$\xi(\sigma)$, such that, by applying a central limit argument,
its fluctuations become
$$
\sqrt{\langle \xi(\sigma)^2 \rangle} = \frac{ \langle
\xi(\sigma)^2 \rangle }{\sqrt{(\beta-1)}} \frac{\beta^3}{3}
$$
which diverge as soon as the denominator approaches zero (i.e. for
$\beta \rightarrow 1^-$).
\newline
Finding the critical exponent happens on the same line by
rewriting eq.(\ref{scaling}) as
$$
\sqrt{\langle m^2 \rangle }\sqrt{(\beta-1)} \sim \langle m^2
\rangle \frac{\beta^3}{3}
$$
and considering, close to criticality, $\beta^3 \sim 1$, which
immediately yields \be \sqrt{\langle m^2 \rangle }  \sim
(\beta-1)^{\frac12}\ee
according to eq.(\ref{prova}).

\begin{remark}  Using eq.(\ref{espa}) to work out an expansion of
the cavity function we obtain

\be\label{expa} \Psi(t) = \int_0^t dt \langle m \rangle_t =
\int_0^t dt \Big( \langle m^2 \rangle t - \langle m^2 \rangle^2
\frac{t^3}{3} + O(\langle m^2 \rangle^4) \Big) \ee
which gives
\be\label{ezpas} \Psi(t) = \langle m^2 \rangle \frac{t^2}{2}  -
\langle m^2 \rangle^2 \frac{t^4}{12} +  O(\langle m^2 \rangle^4)
\ee in perfect agreement with the expansion of the logarithm of
the hyperbolic cosine.
\end{remark}
\textbf{Note} The same method, respectively applied on the SK and
on the Viana-Bray model \cite{viana} of diluted spin glass, has
been discussed in \cite{barra3} and \cite{barra5}.
\begin{remark}
Using the expansion (\ref{ezpas}) for the free energy expression
in Theorem (\ref{215}) we obtain \be \alpha(\beta) = \ln 2 +
\frac{\beta}{2}(\beta-1)\langle m^2 \rangle -
\frac{\beta^4}{12}\langle m^2 \rangle^2 + ... \ee by which we
argue the critical point must be $\beta_c=1$. This can be seen as
follows: Let us note that $A(\beta)=(\beta/2)(\beta-1)$  is the
coefficient of the second order of the expansion in power of the
order parameter (i.e. $\sqrt{\langle m^2 \rangle}$. In the ergodic
phase (with preserved symmetry) the minimum of the free energy
corresponds to a zero order parameter (i.e. $\sqrt{\langle m^2
\rangle} = 0$). This implies that $A(\beta) \geq 0$. Anyway,
immediately below the critical point values of the order parameter
different from zero are possible if and only if $A(\beta) \leq 0$
and consequently at the critical point $A(\beta)$ must be zero.
\newline
This identifies the critical point $\beta_c=1$.
\newline
Coherently, for the same reason the first order term in the
expansion must be identically zero.
\end{remark}
\textbf{Note} An identical approach holds also for the SK spin
glass model \cite{barra1}.

\subsection{Self-averaging properties}

As a sideline, to try and make the work as close as possible to a
guide for more complex models, it is possible to derive the
''locking" of the order parameter, which, in other context (i.e.
spin glasses) is found as a set of equations called
Ghirlanda-Guerra \cite{gg} and Aizenman-Contucci \cite{ac}, while
in simpler systems as the one we are analyzing, not surprisingly
\cite{cg2}, do coincide with just one kind of self-averaging.
\newline
The idea we follow \cite{barra1}\cite{barra2}\cite{barra4} is
deriving filled monomial with respect to the interpolating
parameter, remembering that, in the thermodynamic limit, they do
not depend on such a parameter end evaluating the ''fillable"
result (which do depends on $t$) at $t=\beta$ to free the measure
from the perturbing cavity field.
\begin{proposition}
The self-averaging properties, consequence of the invariance of
filled monomials with respect the perturbing field, hold in the
thermodynamic limit; an example being
\be\label{GGAC} 0 = \lim_{N \rightarrow \infty} \partial_t \langle
m_N^2 \rangle = \langle m^3 \rangle_t - \langle m^2 \rangle
\langle m \rangle_t = \langle m^4 \rangle - \langle m^2 \rangle^2
\ee
\end{proposition}
Even though we followed the derivation presented in \cite{barra1}
(and deepen  in \cite{barra4}  for its dilute variant) to obtain
such constraints, for the Ising model it is straightforward to
check that the original idea presented in \cite{gg} concerning the
self-averaging of the internal energy shares the same relation. In
fact, defining $\langle E \rangle = \lim_{N \rightarrow \infty
}E_N$ and $E_N = H_N(\sigma) / N$, by direct evaluation we have
\begin{remark}
The self-averaging property of the order parameter is a
consequence of self-averaging of the internal energy
$$
\lim_{N \rightarrow \infty} \left( \langle E_N \rangle^2 - \langle
E_N^2\rangle \right) = 0 \Rightarrow \left( \langle m^2 \rangle^2
- \langle m^4 \rangle \right) = 0
$$
\end{remark}
\textbf{Note} In this system without disorder the AC relations and
the GG identities do coincide because of the absence of the
external average over the noise, which introduce different kinds
of self-averaging as discussed for instance in \cite{franz}.
\newline
A less known alternative, richer of surprises, emerges again when
investigating the cavity function. Of course in simple system such
investigation will not tell us much more than what showed so far,
but, remembering we want to show a working method  more than the
results themselves it offers for this particular system, we want
to explore this last variant.
\newline
Remembering Theorem \ref{saturi} and Proposition \ref{busillis}
let us rewrite the free energy according to \be\label{recall}
\alpha(\beta) = \ln 2 + \ln \cosh( t \sqrt{\langle m
\rangle_t}t)\Big|_{t=\beta} - \frac{\beta}{2}\sqrt{\langle m^2
\rangle} \ee
and emphasize that the total derivative with respect to $\beta$ is
\be  \frac{d\alpha(\beta)}{d\beta}=
\frac{\partial\alpha(\beta)}{\partial\beta} +
\frac{\partial\alpha(\beta)}{d\sqrt{\langle m^2
\rangle}}\frac{\partial \sqrt{\langle m^2 \rangle}}{d\beta}. \ee
while, from the general law of thermodynamics \cite{parisi2}, we
know the total derivative of the free energy with respect to
$\beta$ is the internal energy \be \frac{d\alpha(\beta)}{d\beta}=
\frac12 (\sqrt{\langle m^2 \rangle})^2. \ee With this preamble let
us move evaluating the partial derivative of the free energy still
with respect $\beta$:
\begin{eqnarray}\nonumber
\frac{\partial \alpha(\beta)}{\partial \beta} &=&
-\frac{1}{2}(\sqrt{\langle m^2 \rangle})^2 + \Big( \sqrt{\langle m
\rangle_t} \tanh(\sqrt{\langle m \rangle_t}t) \Big)\Big|_{t=\beta}
\\ \nonumber &=& -\frac{1}{2}(\sqrt{\langle m^2 \rangle})^2 + \Big(
\sqrt{\langle m^2 \rangle} \tanh(\sqrt{\langle m^2 \rangle}\beta
\Big)
\end{eqnarray}
which thanks to self-consistency for the order parameter (Eq.
(\ref{tgh})) becomes \be -\frac12(\sqrt{\langle m^2 \rangle})^2 +
(\sqrt{\langle m^2 \rangle})^2 = \frac12 (\sqrt{\langle m^2
\rangle})^2 \ee
hence \be\label{pv1} \frac{\partial\alpha(\beta)}{d\sqrt{\langle
m^2 \rangle}}\frac{\partial \sqrt{\langle m^2
\rangle}}{d\beta}=0.\ee
Let us split the evaluation of Eq. (\ref{pv1}) in two terms
$A$,$B$ (such that the equation reduces to $AB=0$) by defining and
evaluating
\begin{eqnarray}
A &=& \frac{\partial\alpha(\beta)}{d\sqrt{\langle m^2 \rangle}} =
\beta\Big(  \sqrt{\langle m^2 \rangle} - \tanh(\beta \sqrt{\langle
m^2 \rangle}) \Big) \\ B &=& \frac{\partial \sqrt{\langle m^2
\rangle}}{d\beta} = \frac{N}{4\sqrt{\langle m^2 \rangle}}\Big(
\sqrt{\langle m^4 \rangle} - (\sqrt{\langle m^2 \rangle})^2 \Big).
\end{eqnarray}
Putting together the results $AB=0$ we obtain \be \beta\Big(
\sqrt{\langle m^2 \rangle} - \tanh(\beta \sqrt{\langle m^2
\rangle}) \Big)\frac{N}{4\sqrt{\langle m^2 \rangle}}\Big(
\sqrt{\langle m^4 \rangle} - (\sqrt{\langle m^2 \rangle})^2
\Big)=0. \ee This equation acts as a bound and, thought in terms
of the expression (\ref{pv1}), has a vague variational taste. As
in simple system it does not tell us much more than that the
product of self-consistency and self-averaging goes to zero faster
than $N^{-1}$, in complex system has a key role both in defining
the locking of the order parameters \cite{barra1} as in
controlling the system at criticality \cite{barra5}. Furthermore
in such equation the two key ingredient for the behavior of the
system, i.e. self-consistency and self-averaging, appear together
as a whole.

\subsection{Hamilton-Jacobi formalism: order parameter self-averaging and response to field}

This section has been adapted from the work \cite{Gsum} where the
method, in the framework of spin glasses, were originally
developed.
\newline
Next step is investigating the self-averaging of the magnetization
itself. This can be achieved in several ways also within the
interpolating techniques. For the sake of completeness we want to
show a very elegant technique based on two interpolating
parameters.
\newline
\newline
\textbf{The structure of the Hamilton-Jacobi equation}
\newline
\newline
Let us consider a generalized partition function depending on two
parameter $t,x$ (that we are going to think about in terms of
{\itshape generalized time} and {\itshape space}) such that the
corresponding free energy can be written as follows
\be\label{dueF} \alpha_N(t,x)= \frac{1}{N}\ln Z_{N}(t,x) =
\frac{1}{N} \ln \sum_{\sigma} e^{\frac{t}{2N}\sum_{1 \leq i<j \leq
N}\sigma_i\sigma_j + x \sum_{1<i<N}\sigma_i} \ee
and let us consider its $t$ and $x$ streaming (with obvious
meaning, in the averages, of the subscript $\langle \
\rangle_{t,x}$):
\begin{eqnarray} \frac{\partial \alpha_N(t,x)}{\partial t} &=& -\frac{1}{2}
\langle m_N^2 \rangle_{t,x} \\ \frac{\partial
\alpha_N(t,x)}{\partial x} &=&  \langle m_N \rangle_{t,x}
\end{eqnarray}
Let us also define a  $potential$ $V_N(t,x)$ as the variance of
the magnetization in these extended averages:
\be V_N(t,x)= \frac{1}{2}\left( \langle m_N^2 \rangle_{t,x} -
\langle m_N \rangle^2_{t,x} \right)\ee and introduce an Hamilton
function $S_N(t,x)$ as $S_N(t,x)=-\alpha_N(t,x)$. It is now
possible to formulate the next
\begin{proposition} In the generalized space of the interpolants The following Hamilton-Jacobi equation holds
\be\label{HJ} \frac{\partial S_N(t,x)}{\partial t} + \frac{1}{2}
\Big( \frac{\partial S_N(t,x)}{\partial x}\Big)^2 + V_N(t,x) = 0.
\ee
\end{proposition}
The plan now is as follows: Let us try and solve at first the
free-field solution ($V(t,x)=0$), from which the proper solution
of the mean field Ising model (Eq. \ref{freeE}) will follow and we
will argue that $\lim_{N \rightarrow \infty}(\langle m_N^2\rangle
- \langle m_N \rangle^2) = 0$.
\newline
\newline
\textbf{The free field solution: self-averaging}
\newline
\newline
If the $t$-dependent potential is zero then the energy is a
constant of motion such that the $Lagrangian$ $\mathcal{L}$, which
is trivially $\frac{1}{2} \Big( \frac{\partial S_N(t,x)}{\partial
x}\Big)^2$, does not depend on $t$ (in this bridge with classical
mechanics the interpolating parameter $t$ takes the same meaning
of time) and the trajectories of motion are the straight lines
$x(t) = x_0 + \langle m \rangle t$.
\newline
If we denote by a bar the Hamilton function which satisfies the
free-field problem, such solution $\bar{S}(t,x)$ can be worked out
finding a point in the space of solution plus the integral of the
Lagrangian over the time \be\label{lagra} \bar{S}(t,x) =
\bar{S}(t_0,x_0) + \int dt' \mathcal{L}(t',x) \ee Anyway, as we
already stressed, the Lagrangian, in the free-field problem does
not depend on time and the integral inside the Eq. (\ref{lagra})
turns out to be a simple product, furthermore, as initial point
$(t_0,x_0)$ in the plane $(t,x)$ we choose a generic $x_0$ but
$t_0=0$ as this choice enable us to neglect the two body
interaction in the partition function and the problem becomes
straightforward.
\newline
So we have \be\label{HJ2} \frac{\partial \bar{S}_N(t,x)}{\partial
t} + \frac{1}{2} \Big( \frac{\partial \bar{S}_N(t,x)}{\partial
x}\Big)^2 = 0 \ee
on the trajectories $x = x_0 + \langle m \rangle t$. To enforce
now the generalized partition function defined in (\ref{dueF}) to
be the true one of statistical mechanics, remembering that $S(t,x)
= -\alpha(t,x)$ and so $\bar{S}(t,x) = - \bar{\alpha}(t,x)$, we
must evaluate the solution at $t=\beta,x=0$. The solution is
immediate and is
\begin{eqnarray}\nonumber && \bar{S}(t,x)= \bar{S}(0,x_0) + \int dt
\mathcal{L}(t,x) = - \ln 2 - \ln \cosh (\langle m \rangle t) +
\frac{t}{2}\langle m_N^2\rangle \\ && \bar{\alpha}(\beta) =  \ln 2
+ \ln\cosh(\beta \langle m \rangle) - \frac{\beta}{2}\langle m^2
\rangle \end{eqnarray}
which coincides with the solution of the model (Eq. (\ref{freeE}))
assuming that \be \lim_{N \rightarrow \infty}\sqrt{\langle m^2_N
\rangle} = \langle m \rangle \ee
which is is perfect agreement to our request $V(t,x)=0$.
\newline
\newline
\textbf{Response to a field}
\newline
\newline
We understood that, thank to the global gauge symmetry, we can
think at the cavity field both as an added spin of the system as
well as an external perturbation. Once considered the cavity field
$x \sum_i^N \sigma_i$ as a perturbation it may be interesting
asking what the associated observable is for such a field. It is
immediately to check that the observable is the magnetization. \be
\partial_{x}\frac{1}{N}\ln \sum_{\sigma}e^{- t H_N(\sigma)+ x\sum_i^N\sigma_i}|_{t=\beta,x=0} = \langle m_N
\rangle_{t=\beta,x=0}= \langle m_N \rangle \ee While it may still
look unnecessary for the Ising model we stress that the cavity
field naturally puts in evidence the symmetry of the perturbing
field needed to have a projector (a proper ``active'' selector in
the free energy landscape). In fact, it is immediate to think at
the perturbing field as a  magnetic field of strength $x/\beta$ in
some proper units. In complex systems as spin glasses
understanding the right coupling field it is not immediate and
this property can be of precious help as discussed in
\cite{barra6}.

\section{Parisi-like representation}\label{sei}

As a final section, following the early ideas of Guerra
\cite{private}, we try and introduce a formalism close to the
Parisi scheme for spin glasses. This trial is of course not
necessary for the mean field Ising model, but the existence of
this possibility acts as a bridge to a better understanding of the
Parisi theory itself.

\subsection{The order parameter}

Writing equation (\ref{051}) via the cavity function (\ref{052})
we get \be \ln Z_{N+1}(\beta \frac{N+1}{N}) = \ln Z_N(\beta) +
\Psi(\beta) + \ln 2, \ee which can be iterated $N-1$ steps
approaching the recursive relation
\be\label{055} \alpha_{N+1}= \frac{N}{N+1}\ln2 +
\sum_{1<i<N-1}\frac{1}{N+1}\Psi(\beta\frac{N-i}{N+1}) +
\frac{1}{N+1}\alpha_1(\frac{\beta}{N+1}). \ee
Let us take the thermodynamic limit of eq.(\ref{055}): It is
immediate to check that the third term of the r.h.s. goes to zero
while in the second term the summation converges to a Riemann
integral and the first term becomes $\ln2$:

\be\label{056} \lim_{N \rightarrow \infty}\alpha_{N+1}(\beta) =
\alpha(\beta) = \ln 2 + \int_0^1 d\tilde{m} (\beta(1-\tilde{m}))
\ee
being
\be \tilde{m} = \lim_{N\rightarrow \infty} \frac{i}{N}. \ee
Let us now introduce an auxiliary function as

\be \Phi(\tilde{m}) = \ln \langle
e^{f(\tilde{m},\frac{\tilde{m}}{N}\sum_i^N\sigma_i)} \rangle \ee
where in the dependence on $f(\tilde{m},y(\tilde{m}))$ there is
the boundary constraint \be f(1,y) = \ln\cosh(\beta y) \ee such
that \be \Phi(1) = \ln \langle e^{f(1,\frac{1}{N}\sum_i \sigma_i)}
\rangle = \Psi(t=\beta) \ee
Let us look now for the condition under which $\Phi(\tilde{m})$
does not depend on $\tilde{m}$ (i.e. $d_{\tilde{m}} \Phi=0$): for
the sake of convenience, let us introduce
$$
\tilde{f}(\tilde{m}) =
f(\tilde{m},\frac{\tilde{m}}{N}\sum_i\sigma_i), \ \langle a
\rangle_f = \frac{\langle a e^{\tilde{f}} \rangle}{\langle
e^{\tilde{f}} \rangle}
$$
with which we write \be \frac{d\Phi}{d\tilde{m}}=\langle
\partial_{\tilde{m}} \tilde{f} \rangle_f + \frac{1}{N}\sum_i^N \langle
\sigma_i
\partial_y \tilde{f} \rangle_f \ee and let us consider the
following bounds \be |\frac{1}{N}\sum_i^N\langle \sigma_i
\partial_y \tilde{f} \rangle_f| \leq \frac{1}{N}\sum_i^N|\langle
\sigma_i \partial_y \tilde{f} \rangle_f| \leq \frac{1}{N}\sum_i^N
\langle |\partial_y \tilde{f} | \rangle_f = \langle |\partial_y
\tilde{f}| \rangle_f \ee which allow one to introduce a function
$x:[0,1]\rightarrow[-1,+1]$ such that \be\label{modulus}
\frac{1}{N}\sum_i^N \langle \sigma_i \partial_y \tilde{f}
\rangle_f = x(\tilde{m}) \langle |\partial_{\tilde{m}} \tilde{f}|
\rangle_f \ee
\begin{remark}
The existence of the function {\itshape modulus} inside the r.h.s.
of Eq. (\ref{modulus}) allows one to take into account just one
branch at time with complete symmetry between the two branches.
This reflects the properties of the magnetization in the broken
ergodicity phase.
\end{remark}
with the scope of moving the independence condition of $\Phi$ from
$\tilde{m}$ in the choice of $f$, which must obey the following
differential problem:

\subsection{The Parisi-like equation}
\be\left\{ \begin{array}{l} \label{parisi}
 \partial_{\tilde{m}} f(\tilde{m},y)+x(\tilde{m})|\partial_y f(\tilde{m},y)|  = 0 \\
 f(1,y) = \ln \cosh (\beta y)
\end{array}\right.\ee
and remember that $f(0,0)= \Phi(0) = \Phi(1) = \Psi(t=\beta) $.
\newline
\begin{remark}
The above equation immediately reveals a big difference between
the Ising model and the SK: linearity. In fact the Parisi equation
for the spin glasses \cite{MPV} is non linear and shows several
bifurcation points, while, in the problem (\ref{parisi}), once
chosen a branch, the evolution is unique.
\end{remark}
To start solving (\ref{parisi}) let us switch to a $p$ variable
such that \be p = - \int_{\tilde{m}}^{1} d\tilde{m}'x(\tilde{m}')
\ee by which the Parisi-like equation for the Ising model turns
out to be solvable with the D'Alamber technique. Calling in fact
$\tilde{m} \rightarrow p \Rightarrow   f(\tilde{m}(p),y)
\rightarrow g(p,y)$ we get
$$
\partial_p g(p,y) + \partial_y g(p,y) = 0
$$
solved by $g(p,y)= \ln\cosh(t(p+y)) \rightarrow f(q,y) = \ln\cosh(t(y \pm \int_q^1 dq'x(q')))$,
where the $\pm$ signs are choosen accordingly to the branch of the choosen derivative of $f$ with respect to $y$.
\newline
Solving for the $\Psi(t=\beta)$ we get

\be\label{comparison} \Psi(t=\beta) = \ln\cosh(\beta \int_0^1
d\tilde{m}'x(\tilde{m}')) \ee
\textbf{Comparison of the order parameters}
\newline
\newline
Let us now equate Eq. (\ref{compegno}) with Eq.
(\ref{comparison}): We immediately obtain
\be \sqrt{\langle m^2 \rangle} = \langle m \rangle = \int_0^1
d\tilde{m} x(\tilde{m}) \ee
by which we argue that the function $x(\tilde{m})$ has the meaning
of a probability density for the order parameter (i.e. the
magnetization). Further one could go beyond this scheme, but this
will not be discussed here, working out the equivalent of the
broken replica bound to make sharper statements concerning the
$x(\tilde{m})$ following \cite{broken}.
\begin{remark}
Another possibility is by exploring the {\itshape replica trick}
method \cite{MPV} assigning a delta-like probability distribution
for the interaction matrix $J_{ij}$ (i.e. $P(J_{ij}) \sim
\delta(J_{ij}-1)$) which factorizes replicas and no ansatz is
required in this simple case.
\end{remark}

\section{Conclusion}

In this paper we have studied the mean field Ising model with the
interpolating techniques. These methods, which have been at the
basis of a recent breakthrough in spin glass theory turn out to be
of great generality, property that has been successfully tested
investigating this simpler model. Several techniques, linked one
another by the interpolation method, have been shown throughout
the paper: key ingredients for the free energy thermodynamic limit
are the sub-additivity and the bounds in the volume size. Another
central role is played by the gauge invariance when analyzing the
expression of the free energy itself: via this symmetry the cavity
field becomes a perturbing external field (what is called
{\itshape stochastic stability} in spin glass literature) and
viceversa and the synergy between the two approaches enables one
to work out several properties of the model as the critical
behavior and the self-averaging relations. The technique with two
interpolating parameters has also been discussed: a suitable
streaming of a generalized free energy with respect to these
parameters can bring to the formulation of an Hamilton-Jacobi
equation in the interpolation space by which again the solution of
the model and the self-averaging can be deduced. At the end a
formulation of the theory in terms of Parisi representation is
tried, with particular emphasis on the meaning of the order
parameter.
\newline
As a last remark we stress that this work has been written with
the aim of developing a simpler but dense exercise of statistical
mechanics to make these techniques ready to be used to the reader
not familiar with the field of spin-glasses.

\section{Acknowledgment}
The author is pleased to thank Francesco Guerra for a priceless
scientific interchange and Pierluigi Contucci for several useful
conversations. This work is partially supported by the MIUR within
the Smart-Life Project (Ministry Decree $13/03/2007$ n.$368$) and
partially by a King's College London grant.


\addcontentsline{toc}{chapter}{References}
\begin{thebibliography}{9}

\bibitem{amit} D.J. Amit, {\em Modeling brain function: The world of attractor neural
network} \ Cambridge Univerisity Press, (1992)

\bibitem{barra3} A. Agostini, A. Barra, L- De Sanctis
\textit{Positive-Overlap transition and Critical Exponents in mean
field spin glasses}, J. Stat. Mech. P11015 (2006).

\bibitem{ac} M. Aizenman, P. Contucci, {\em On the stability of the
quenched state in mean field spin glass models}, J. Stat. Phys. {\bf 92},
765-783 (1998).

\bibitem{ass} M. Aizenman, R. Sims, S. L. Starr,
\emph{An Extended Variational Principle for the SK Spin-Glass
Model}, Phys. Rev. B, \textbf{68}, 214403 (2003)

\bibitem{barabasi}  Reka Albert, Albert-Laszlo Barabasi,
\textit{ Statistical mechanics of complex networks},  Reviews of
Modern Physics \textbf{74}, 47 (2002)

\bibitem{barra1} A. Barra,
\textit{Irreducible free energy expansion and overlap locking in
mean field spin glasses}, J. Stat. Phys. \textbf{123}, 601-614
(2006).

\bibitem{barra2} A. Barra, L- De Sanctis
\textit{Overlap fluctuation from Boltzmann random overlap
structure}, J. Math. Phys. \textbf{47}, 103305 (2006).


\bibitem{barra4} A. Barra, L- De Sanctis
\textit{Stability Properties and probability distributions of
multi-overlaps in diluted spin glasses}, J. Stat. Mech. P08025
(2007).

\bibitem{barra6} A. Barra, L- De Sanctis
\textit{Spin-Glass transition as the failure of saturability}, To
appear (2007).

\bibitem{barra5} A. Barra, L- De Sanctis, V. Folli
\textit{Critical behavior of random spin systems},  {\tt
cond-mat/0710.4472}

\bibitem{boviergrem} A. Bovier, I. Kurkova, {\em
Rigorous results on some simple spin glass models}, {\tt
cond-mat/0206562}.

\bibitem{bovier} A. Bovier, I. Kurkova, M. L\"owe,
{\em Fluctuations of the free energy in the REM and the $p$-spin
SK models}, to appear on Ann. Probab.

\bibitem{comets} F. Comets, J. Neveu, {\em The Sherrington-Kirkpatrick
model of spin glasses and stochastic calculus: the high
temperature case}, Commun. Math. Phys. {\bf 166}, 549 (1995).

\bibitem{ton3} A.C.C. Coolen {\em The Trick Which Became a Theory: A Brief History of the Replica Method}
available at http://www.mth.kcl.ac.uk/~tcoolen/

\bibitem{contu1} P. Contucci, S. Ghirlanda, {\em Modeling Society with Statistical Mechanics: an Application to
Cultural Contact and Immigration},  Quality and Quantity, Vol.
\textit{41}, 569-578 (2007)

\bibitem{cg2} P. Contucci, C. Giardin\`a,
{\em Spin-Glass Stochastic Stability: a Rigorous Proof},
\texttt{math-ph/0408002}.

\bibitem{contu2} P. Contucci, J. Lebowitz {\em Correlation Inequalities for Spin Glasses}
        To appear in: Annales Henri Poincare, \texttt{cond-mat/0612371}.

\bibitem{luca} L. De Sanctis {\em General Structures for Spherical and Other Mean-Field Spin Models},  J. Stat. Phys. \textbf{126}

\bibitem{franz} L. De Sanctis. S. Franz {\em Self averaging identities for random spin
systems}  \texttt{math-ph/0705:2978}.

\bibitem{ellis} R.S. Ellis,
{\em Large deviations and statistical mechanics}, Springer, New
York (1985).

\bibitem{nota} a key ingredient for the existence of the
thermodynamic limit is  the subadditivity or to the
superadditivity of the free energy with respect the system size.
The Hopfield model shows, varying the storaged memory, both the
features. As a consequence there is a region of unknown width in
which the model free energy is nor subadditive neither
superadditive.


\bibitem{gg} S. Ghirlanda, F. Guerra, {\em General properties of overlap
distributions in disordered spin systems. Towards Parisi ultrametricity},
J. Phys. A, {\bf 31} 9149-9155 (1998).

\bibitem{fischer} K. H. Fischer, J. A. Hertz, {\em Spin glasses}, Cambridge
University Press, Cambridge (1991).

\bibitem{gallo} I. Gallo, P. Contucci, {\em Bipartite Mean Field Spin System: Existence and
Solution}, ArXiv:cond-mat/0710.0800

\bibitem{g2} F. Guerra, \emph{Mathematical Aspects of Mean Field Spin Glass Theory},
 \texttt{cond-mat/0410435}.

\bibitem{g} F. Guerra, \emph{About the Cavity Fields in Mean Field Spin Glass Models},
 \texttt{cond-mat/0307673}.

\bibitem{Glocarno} F. Guerra, {\em Fluctuations and thermodynamic variables
in mean field spin glass models}, in ``Stochastic provesses,
physics and geometry, II'', S. Albeverio {\em et al.} eds.,
Singapore (1995).

\bibitem{broken} F. Guerra, {\em Broken Replica Symmetry Bounds in the
Mean Field Spin Glass Model}, Commun, Math. Phys. \textbf{233:1},
1-12 (2003)

\bibitem{guerra2} F. Guerra, {\em About the overlap distribution in mean field
spin glass models}, Int. Jou. Mod. Phys. B {\bf 10}, 1675-1684
(1996).

\bibitem{guerra4} F. Guerra, {\em The cavity method in the mean field
spin glass model. Functional reperesentations of thermodynamic
variables}, in ``Advances in dynamical systems and quantum
physics'', S. Albeverio {\em et al.} eds., Singapore (1995).

\bibitem{Gsum} F. Guerra, {\em Sum rules for the free energy in the mean
field spin glass model}, in {\em Mathematical Physics in
Mathematics and Physics: Quantum and Operator Algebraic Aspects},
Fields Institute Communications {\bf 30}, American Mathematical
Society (2001).

\bibitem{private} F. Guerra, private communications.

\bibitem{leshouches} F. Guerra, {em An introduction to mean field
spin glass theory: methods and results}, Lecture at Les Houches
winter school (2005)

\bibitem{limterm} F. Guerra, F. L. Toninelli, {\em
The Thermodynamic Limit in Mean Field Spin Glass Models}, Commun.
Math. Phys. {\bf 230:1}, 71-79 (2002).

\bibitem{gt2} F. Guerra, F. L. Toninelli,
\emph{The high temperature region of the Viana-Bray diluted spin
glass model}, J. Stat. Phys. \textbf{115} (2004).

\bibitem{CLT} F. Guerra, F. L. Toninelli, {\em Central limit theorem for
fluctuations in the high temperature region of the
Sherrington-Kirkpatrick spin glass model}, J. Math. Phys., to
appear, {\tt cond-mat/0201092}.

\bibitem{limterm2} F. Guerra, F. L. Toninelli, {\em
The infinite volume limit in generalized mean field disordered
models}, Markov Processes and Rel. Fields, to appear, {\tt
cond-mat/0208579}.

\bibitem{ksat}  Stephan Mertens, Marc Mezard, Riccardo Zecchina,
    {\em Threshold values of Random K-SAT from the cavity
    method},  to appear in Random Structures and Algorithms,  {\tt
   cond-mat/0309020}.

\bibitem{MPV} M. M\'ezard, G. Parisi and M. A. Virasoro, {\em Spin glass theory
and beyond}, World Scientific, Singapore (1987).

\bibitem{mezard} M. M\'ezard, G. Parisi, N. Sourlas, G. Toulouse, M. A.
Virasoro, {\em Replica symmetry breaking and ultrametricity}, J.
Physique 45, 843 (1984).

\bibitem{pagna} A. Pagnani, G. Parisi, F. Ricci-Tersenghi
{\em Glassy transition in a disordered model for the RNA secondary
structure }, Phys. Rev. Lett. \textit{84}, 2026 (2000)

\bibitem{parisiSS} G. Parisi, {\em Stochastic Stability},
Proceedings of the Conference Disordered and Complex Systems,
London 2000

\bibitem{parisi2} G. Parisi, {\em Statistical field theory} Addison Wesley, New York, 1988

\bibitem{pastur} L. Pastur, M. Shcherbina, {\em The absence of self-averaging
of the order parameter in the Sherrington-Kirkpatrick model}, J.
Stat. Phys. {\bf 62}, 1 (1991).

\bibitem{T} M. Talagrand,
{\em Spin glasses: a challenge for mathematicians. Cavity and Mean
field models}, Springer Verlag (2003).

\bibitem{tala} M. Talagrand, {\em The Parisi formula},  Annals of Mathematics 163, no 1, 2006, 221-263

\bibitem{viana} L. Viana, A. J. Bray, {\em Phase diagrams for dilute
spin-glasses}, J. Phys. C {\bf 18}, 3037 (1985).


\end{thebibliography}
\end{document}